\documentclass[conference,10pt]{IEEEtran}
\IEEEoverridecommandlockouts
\usepackage{cite}
\usepackage{amsmath,amssymb,amsfonts}
\usepackage{algorithmic}
\usepackage{graphicx}
\usepackage{textcomp}
\usepackage{xcolor}
\usepackage{multirow}
\usepackage{array}
\newcolumntype{L}{>{\raggedright\arraybackslash}m{2.1cm}}
\newcolumntype{E}{>{\raggedright\arraybackslash}m{1.5cm}}

\def\BibTeX{{\rm B\kern-.05em{\sc i\kern-.025em b}\kern-.08em
    T\kern-.1667em\lower.7ex\hbox{E}\kern-.125emX}}
\usepackage[letterpaper, portrait, margin=0.7in, bmargin=0.95in]{geometry}

\usepackage{fancyhdr}
\fancypagestyle{firststyle}{\fancyhf{}
	\fancyhead[L]{D. Shakya, T. Wu, M. E. Knox and T. S. Rappaport, "A Wideband Sliding Correlation Channel Sounder in 65 nm CMOS: Evaluation Board Performance," in \textit{IEEE Transactions on Circuits and Systems II: Express Briefs, } Sep. 2021, pp. 1--5}
}

\begin{document}

\title{A Wideband Sliding Correlation Channel Sounder in 65 nm CMOS: Evaluation Board Performance\\
}

\author{\IEEEauthorblockN{Dipankar Shakya$^{1}$, Ting Wu$^{2}$, Michael E. Knox$^{1}$, and Theodore S. Rappaport$^{1}$}
	\IEEEauthorblockA{\textit{NYU WIRELESS$^{1}$ and Center for Neural Science$^{2}$, New York University, USA}\\
		\{dshakya, ting.wu, mikeknox, tsr\}@nyu.edu}
}

\maketitle

\linespread{1.05}

\thispagestyle{firststyle}

\begin{abstract}
Emerging applications such as wireless sensing, position location, robotics, and many more are driven by the ultra-wide bandwidths available at millimeter-wave (mmWave) and Terahertz (THz) frequencies. The characterization and efficient utilization of wireless channels at these extremely high frequencies require detailed knowledge of the radio propagation characteristics of the channels. Such knowledge is developed through empirical observations of operating conditions using wireless transceivers that measure the impulse response through channel sounding. Today, cutting-edge channel sounders rely on several bulky RF hardware components with complicated interconnections, large parasitics, and sub-GHz RF bandwidth. This paper presents a compact sliding correlation-based channel sounder baseband built on a monolithic integrated circuit (IC) using 65 nm CMOS, implemented as an evaluation board achieving a 2 GHz RF bandwidth. The IC is the world's first gigabit-per-second channel sounder baseband implemented in low-cost CMOS. \textcolor{black}{The presented single-board system can be employed at both the transmit and receive baseband to study multipath characteristics and path loss.} Thus, the single-board implementation provides an inexpensive and compact solution for sliding correlation-based channel sounding with 1 ns multipath delay resolution.         
\end{abstract}

\begin{IEEEkeywords}
142 GHz, Channel sounder, mmWave, On-chip baseband, PN sequence, RF hardware, Sliding correlation, THz, XPD 
\end{IEEEkeywords}

\section{Introduction}  
%*applications
%*millimeter wave ICs and channel sounding
Modern silicon integrated circuit (IC) technology has become capable of fabricating components for wireless communication at Terahertz (THz) frequencies\cite{KO2019cm}. The current fifth generation (5G) of cellular wireless has adopted millimeter-wave(mmWave) frequencies for operation; sixth generation (6G) cellular wireless is envisaged to leverage the THz frequency range and the wide bandwidths therein for diverse applications, such as augmented/virtual reality, sub-centimeter position location, environmental imaging and sensing, robotics, and cloud computing\cite{Rappaport2019ia,ghosh2019ia}. The short wavelengths and higher attenuation in mmWave and THz frequency bands necessitate ultra-wideband, highly sensitive, and noise-immune hardware for the study, measurement, and accurate modeling of the wireless channels at these frequencies for their effective utilization \cite{Akyildiz2014pc}.             

The hardware currently used for wireless channel sounding involves using sophisticated RF equipment or multiple commercial-off-the-shelf (COTS) components from different manufacturers\cite{pirkl2008wc,Wu2019iscas,Mac2017sac}. A COTS system-based sliding correlation channel sounding involves complicated integration of several bulky and expensive test equipment with numerous fragile cable connections\cite{Wu2019iscas}. Most COTS sliding correlator systems are also limited in the maximum baseband chip rate and achieve RF bandwidth under 1 GHz\cite{Mac2017sac,Zhang2018icc}. \textcolor{black}{Thus, there is motivation for a compact integrated channel sounder system to supersede COTS systems for resolution of multipath components in mmWave and THz channels crucial for enabling diverse applications envisioned in the future\cite{Rappaport2019ia,Xing2021cl}.}        

\textcolor{black}{This paper presents the performance and implementation results of an evaluation board (EVB) fabricated for a monolithic channel sounder integrated circuit (IC) in 65 nm CMOS that is ultra-wideband, programmable, and synchronization-capable. The EVB enables integration of the IC with a sliding correlation channel sounder, replacing the entire system baseband with a single board. The paper is organized as follows: Section II highlights the fundamental concepts behind the channel sounder IC design and intended implementation. Section III details the EVB design fabrication. Section IV presents the measurement results that verify the EVB performance with theorized output. Next, Section V highlights the EVB integration with a 142 GHz channel sounder. Then, Section VI presents the measurement of antenna cross-polarization effect using the EVB integrated system.}  

\section{Design of the on-chip sliding correlator based channel sounder and EVB}\label{sec2}
Channel sounding is an experimental means of characterizing the wireless channel between a transmitter (TX) and receiver (RX). Sliding correlation based channel sounding allows studying wireless signal propagation in the time domain using autocorrelation between a faster pseudorandom noise (PN) sequence at chip-rate equal to a fast-clock frequency $\alpha$, and a slower replica sequence at rate equal to slow-clock $\beta$. The sliding correlation method introduces a time-dilation of the processed signal by a slide factor $\gamma$, given in \eqref{gama}, that results in a bandwidth compression, and processing gain while rejecting external interference\cite{Wu2019iscas,pirkl2008wc}.

\begin{figure}[htbp]
	\centering
	\includegraphics[width=0.46\textwidth]{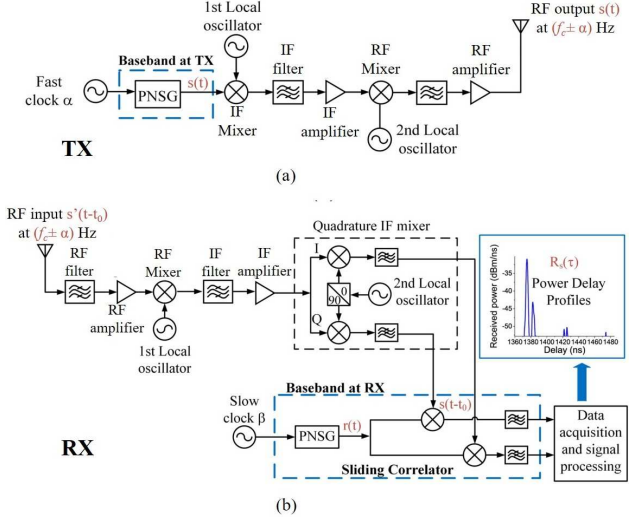}
	\caption{(a) TX system block diagram that generates a PN sequence at chip-rate $\alpha$. (b) RX system block diagram of a sliding correlation channel sounder that generates time dilated power delay profiles, $R_s(\tau)$\cite{Wu2019iscas}. \newline}
	\label{fig:Sl_corr}
	\vspace{-2 em}
\end{figure}

The block diagrams of the TX and RX in a traditional sliding correlation channel sounder system are illustrated in Fig. \ref{fig:Sl_corr}(a) and (b). The PN sequence generator (PNSG) at the TX baseband in Fig. \ref{fig:Sl_corr}(a) generates a PN sequence, $s(t)$, at chip-rate equal to $\alpha$. $s(t)$ is ultimately transmitted at RF frequency $f_{c}\pm\alpha$. Replicas of the signal arrive at the RX over multiple paths with different delays, modified as per the channel's response to the signal impetus. The RX intermediate frequency (IF) mixer, demodulates the received signal into in-phase (I) and quadrature-phase (Q) components, which are mixed with a replica of the transmitted PN sequence, $r(t)$, at rate of $\beta$ to accomplish sliding correlation\cite{Mac2017sac}. The result is a time-dilated power delay profile (PDP) representing the wireless channel response, obtained in \eqref{s_corr_op}. This continuous-time operation can be represented as matrix multiplications between two row vectors $\bar{s}$ and $\bar{r}$ with discrete samples of $s(t-t_{0})$ and $r(t)$ respectively, for each sample delay, $k$, resulting in a vector $\bar{R}$ with sample size ($n$) equal to $\bar{s}$, as shown in \eqref{s_corr_ds}\cite{newhall1997wideband}.
\begin{equation}\label{gama}
\begin{split}
\gamma&= \frac{\alpha}{\alpha-\beta} .\\
\end{split}
\end{equation}

\begin{equation}\label{s_corr_op}
\begin{split}
R_{s}(\tau)&= \int_{0}^{T} r(t)s(t-t_{0}-\tau)dt .\\
\end{split}
\end{equation}

\begin{equation}\label{s_corr_ds}
\begin{split}
R_{k}&= \bar{s}_{(k)}\bar{r}^{T} ; k=1 \text{ to } n .\\
\end{split}
\end{equation}

The path loss (PL) is a fundamental channel property captured by a PDP and may be quantified using the transmit power ($P_{tx}$) fed to a TX-antenna with gain ($G_{tx}$) and the received power ($P_{rx}$) at a distance ($d$) after the RX-antenna with gain ($G_{rx}$), as shown in \eqref{eq:pl}:
\begin{equation}\label{eq:pl}
\begin{split}
PL(d)[dB]= &P_{tx}[dBm] - P_{rx}(d)[dBm]\\ &+ G_{tx}[dBi] + G_{rx}[dBi] .\\
\end{split}
\end{equation}

In addition to experiencing losses, the signal propagation along different paths in the channel may affect the signal polarization and transmitted energy may appear in an orthogonal orientation of the RX-antenna\cite{xing2018vtc,Mac2015ia}. The cross-polarization discrimination (XPD) characterizes this aspect of propagation as the ratio of power received (difference in dB) when TX and RX-antennas are co-polarized to when they are cross-polarized, as stated in \eqref{eq:xpd}. The vertical (V) or horizontal (H) antenna orientations commonly describes co-polarization
(e.g., V-V) or cross-polarization (e.g., V-H)\cite{Wong2017tcas}. 
\begin{equation}\label{eq:xpd}
\begin{split}
XPD(d)[dB]= &PL_{V-V}(d)[dB] - PL_{V-H}(d)[dB] .\\
\end{split}
\end{equation}          
\color{black}
    
\subsection{Design of the Channel Sounder Baseband IC}
\begin{figure}[htbp]
	\centering
	\includegraphics[width=0.46\textwidth]{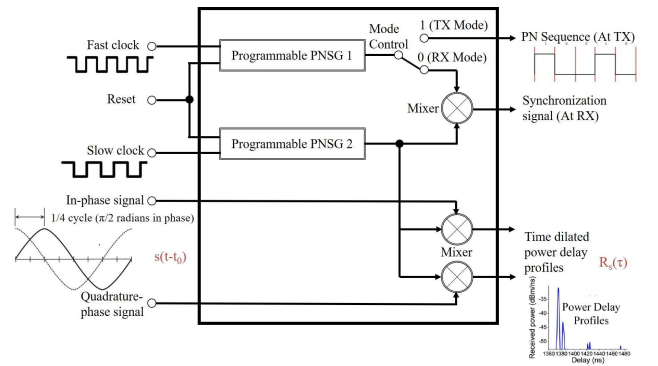}
	\caption{Block diagram of the baseband at TX and RX of the channel sounder on a single IC incorporating two PNSGs, sliding correlator, and \textit{Sync} signal mixer\cite{Shakya2020gc}.\newline}
	\label{fig:slc_ic}
	\vspace{-1.2 em}
\end{figure}

The channel sounder IC, shown in Fig. \ref{fig:slc_ic} and schematics detailed in \cite{Wu2019iscas}, is fabricated in 65 nm CMOS as a monolithic IC and integrates the TX and RX baseband components of the sliding correlation channel sounder system (dashed blue boxes in Fig. \ref{fig:Sl_corr}(a) and (b)). A `Mode Control' pin on the IC switches modes of operation as TX for the pin set `Low' or RX for `High'.

In TX mode, PNSG1 generates a PN sequence of a programmed length at rate equal to $\alpha$ that is directly fed out. In RX mode, the PNSG1 output is mixed with a slower replica generated by PNSG2 at chip-rate of $\beta$ to produce the autocorrelation output as the synchronization (\textit{Sync}) signal. The \textit{Sync} signal is a train of uniformly spaced pulses that may be used as timing reference for aligning the generated PDPs and calculating the absolute delay of multipath components\cite{newhall1997wideband}. The period between pulses, $T_{sync}$, is the duration of the PN sequence dilated by $\gamma$ and is obtained using \eqref{sync_dur}.
 
The output of PNSG2 is also fed in parallel to two other mixers where these are correlated with the demodulated I and Q signals\cite{Wu2019iscas}. The sliding correlation output of the two mixers provides the I and Q components of the time dilated PDP, $R_{s}(\tau)$ in \eqref{s_corr_op}.
\begin{equation}\label{sync_dur}
\begin{split}
T_{sync}[sec]&= \frac{PN\;length}{\alpha}\times\gamma .\\
\end{split}
\end{equation}      

\section{The Fabricated EVB}\label{sec3}
\begin{figure}[htbp]
	\centering
	\includegraphics[width=0.44\textwidth]{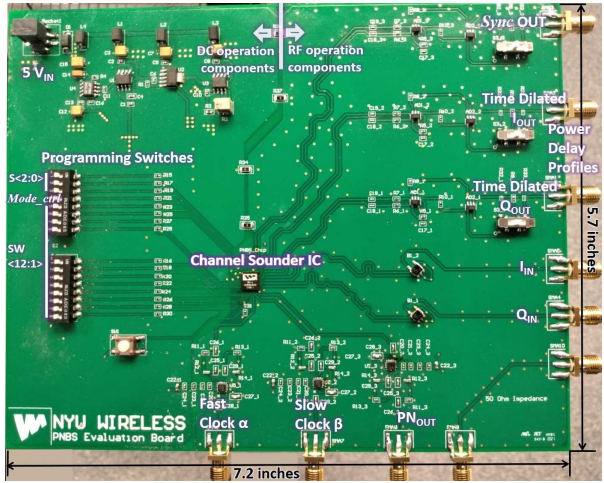}
	\caption{Top view of the sliding correlator based channel sounder EVB with all components assembled\cite{Shakya2020gc}. Channel sounder IC at the center.}
	\label{fig:fin_brd}
	\vspace{-0.5 em}
\end{figure}

The EVB design creates an interfacing environment for the channel sounder IC with traditional sliding correlator based channel sounder setups replacing the TX/RX baseband, such as systems described in \cite{Mac2017sac,Xing2018gc,Zhang2018icc,zwick2005vt}. In particular, the EVB was designed to replace the PNSG, mixers, filters, and amplifiers in the baseband of the 142 GHz channel sounder at NYU WIRELESS\cite{Xing2018gc}. The EVB design specifics and considerations are detailed in \cite{Shakya2020gc}, resulting in the finalized EVB presented in Fig. \ref{fig:fin_brd}.

\color{black}
In order to reduce interference and unwanted coupling, the board was designed to segregate analog signal components on the right side of the board and low frequency elements on the left, as shown in Fig. \ref{fig:fin_brd}.  Analog signal transmission occurs over interconnecting microstrip lines impedance matched to 50$\Omega$ surrounded with grounded vias to maximize signal power across components, reduce fringing fields, and improve isolation between analog traces\cite{Shakya2020gc,Girlando1999tcas}.

Three digital words make up the programming switches: $S\langle2:0\rangle$ for the PN sequence maximal length ($2^{N}-1$, N={5,6,..,12}), \textit{Mode\_Ctrl} to switch between TX and RX operations, and $SW\langle12:1\rangle$ to configure the PNSG and determine the runs of consecutive ``ones" and ``zeros" in the PN sequence. The $SW\langle12:1\rangle$ selects the feedback lines for the 12-stage linear feedback shift register(LFSR) in the PNSG\cite{pirkl2008wc,Wu2019iscas}.

\section{Performance Results}\label{Sec4}
\color{black}
\subsection{TX Mode}
\begin{figure}[htbp]
	\centering
	\includegraphics[width=0.43\textwidth]{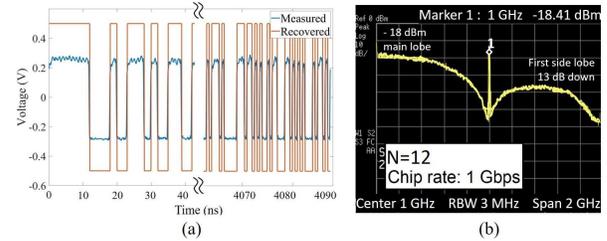}
	\caption{(a) Time domain measurement of the generated PN sequence; analog waveform (blue), chip sequence (orange). (b) Measured spectrum for the PN sequence operating at 1 Gbps for sequence length $N=12$.}
	\label{fig:msmts}
	\vspace{-0.4em}
\end{figure}

As elucidated in Fig. \ref{fig:Sl_corr}(a), PNSG 1 operating using $\alpha=1$ GHz was leveraged to generate a maximal length PN sequence to be transmitted. Setting the programming switches, a 4095 chip PN sequence ($N= 12$; $S\langle2:0\rangle=$ “$111$”)  was generated by the PNSG at a 1 Gigachips-per-second (Gcps) rate, the maximum rate achieved for the IC, as shown in Fig. \ref{fig:msmts}(a)\cite{Wu2019iscas}. The PN sequence output was programmed with SW$\langle12:1\rangle=$ “$000000101001$” for feedback to the [12,6,4,1] stages of the PNSG LFSR. \textcolor{black}{The measured spectrum of the resultant PN sequence showed the expected $sinc^{2}(x)$ profile with distinct null at 1 GHz and first side-lobe 13 dB below the main-lobe, indicating a null-to-null RF bandwidth of 2 GHz, as seen in Fig. \ref{fig:msmts}(b).}

\subsection{RX Mode}
\begin{figure}[htbp]
	\centering
	\includegraphics[width=0.39\textwidth]{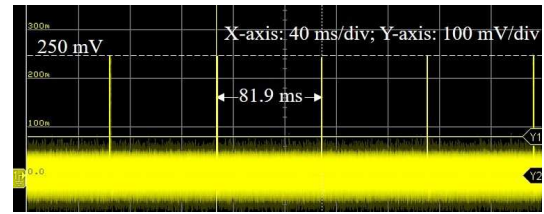}
	\caption{250 mV synchronization peaks resulting from sliding correlation between 1 Gcps and 999.95 Mcps PN sequences. Peaks are separated by the sliding correlation period of  81.9 ms.}
	\label{fig:sync_sig}
	\vspace{-0.5em}
\end{figure}

Setting the ``\textit{Mode\_control}” switch to `$0$’ selected the RX operation. Sliding correlation performed between a maximal length PN sequence at 1 Gcps and a slightly slower replica at 999.95 Mcps followed by low-pass filtering the signal at 100 kHz cutoff resulted in distinct periodic \textit{Sync} signal peaks of ~250 mV that provide a timing reference for PDPs. For a $\gamma$ equal to 20000 and 4095 chip PN sequence at 1 Gcps, from \eqref{sync_dur} a sliding correlation period of, $T_{sync}=$ 81.9 ms was observed between peaks, as in Fig. \ref{fig:sync_sig}. 

\begin{figure}[htbp]
	\centering
	\includegraphics[width=0.48\textwidth]{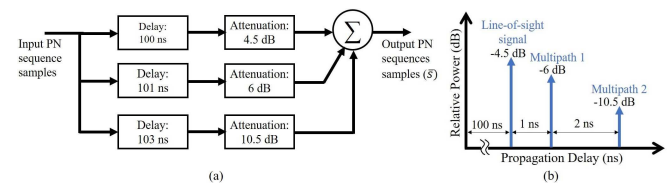}
	\caption{(a) Replicas of the transmitted PN sequence delayed, attenuated, and added to emulate the demodulated signal at RX ($\bar{s}$). (b) Resulting channel response with LOS and multipath signals separated by nanosecond intervals.}
	\label{fig:RX_sig}
	\vspace{-0.5em}
\end{figure}

The RX baseband sliding correlation was also tested for a 1 ns multipath delay resolution using the PN sequence replica, at 999.95 Mcps, and emulated I and Q components of the demodulated PN sequence fed in as EVB inputs. For testing, the demodulated I signal was generated using an arbitrary waveform generator (AWG). First, a 2047 chip (N=11) PN sequence output at 1 Gcps was generated and captured as discrete samples. Then, three replicas of the sequence were added sample-by-sample with delays of 1 ns, and 2 ns in between and attenuation of 4.5 dB, 6 dB, and 10.5 dB respectively resulting in, $\bar{s}$, as shown in Fig. \ref{fig:RX_sig}(a). Thus, a scenario of a line-of-sight (LOS) signal and two multipath components at the RX was simulated, as shown in Fig. \ref{fig:RX_sig}(b).
\vspace{-0.5em}

\begin{figure}[htbp]
	\centering
	\includegraphics[width=0.44\textwidth]{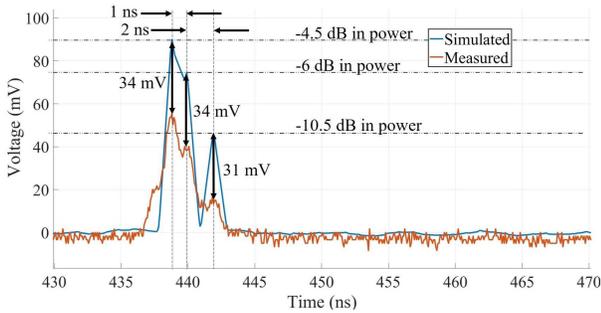}
	\caption{Peak comparison of recorded and undilated PDP (orange) from the EVB with mathematically generated output (blue). 1 ns separation between first two peaks verify theorized multipath delay resolution. Amplitude offset due to AWG output voltage limitation and cable losses.}
	\label{fig:PDP}
	\vspace{-0.5em}
\end{figure}

$\bar{s}$ was then loaded onto the AWG and an approximate $s(t-t_{0})$ was fed into the EVB for sliding correlation as defined by \eqref{s_corr_op}. The final result was the PDP representing the time-domain response of the wireless channel, corresponding to the scenario in Fig. \ref{fig:RX_sig}(b). Samples of the slower PN sequence at 999.95 Mcps were also captured from the oscilloscope as $\bar{r}$ to simulate the resulting PDP using \eqref{s_corr_ds}. The measured PDP was recorded, undilated, and compared to $\bar{R}$, as presented in Fig. \ref{fig:PDP}. The PDP amplitude was observed lower than that simulated as the AWG output was limited to $\pm250$ mV while the EVB generated output at $\pm300$ mV. Further, the PDP may also be time-aligned based on the \textit{Sync} signal to match the 100 ns delay of the first LOS component.   

In comparison to COTS systems described in \cite{Mac2017sac}, the EVB provides a finer multipath resolution of 1 ns on a single-board system, as shown in Fig. \ref{fig:PDP}. The channel sounder EVB offers distinct advantages in size, weight, and simplicity compared to setups that achieve a finer multipath resolution ($<$ 1 ns) over hardware\cite{Xing2018gc,Takeuchi2001pimrc}, or software via frequency concatenation\cite{zwick2005vt}. A general comparison between a sample set of sliding correlation channel sounders is presented in Table.\ref{tab:specs}.
\vspace{-1.0em}

\begin{table}[htbp]
	\caption{Comparison of channel sounder hardware specifications}
	\label{tab:specs}
	\begin{center}
		\begin{tabular}{|L|E|E|E|}
			\hline
			{\bf Channel sounder specifications} & {\bf NYU WIRELESS\cite{Xing2018gc}} & {\bf US Naval Academy\cite{Zhang2018icc}} & {\bf NYU WIRELESS EVB\cite{Shakya2020gc}} \\
			\hline
			{\bf Carrier frequency} &    142 GHz &     28 GHz &    142 GHz \\
			\hline
			{\bf Component type} &       COTS &       COTS & COTS+EVB \\
			\hline
			{\bf Monolithic IC integration} &         No &         No & Yes (65 nm CMOS) \\
			\hline
			{\bf Max chip rate} &     2 Gcps &   0.4 Gcps &     1 Gcps \\
			\hline
			{\bf RF bandwidth} &      4 GHz &    800 MHz &      2 GHz \\
			\hline
			{\bf Multipath delay resolution} &     0.5 ns &     2.5 ns &       1 ns \\
			\hline
			{\bf Synchronization} & External: Rubidium clock & External: GPS disciplined & On-board support: \textit{Sync} signal \\
			\hline
		\end{tabular}  
	\end{center}
	\textit{GPS: Global Positioning System}
	\vspace{-1.5 em}
\end{table}

\color{black}
\section{Integration with the Channel Sounder System}
\begin{figure}[htbp]
	\centering
	\includegraphics[width=0.44\textwidth]{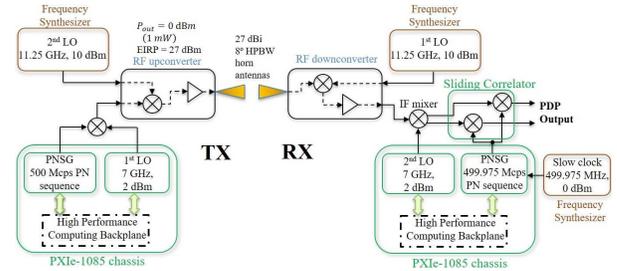}
	\caption{Simplified block diagram representation of the 142 GHz channel sounding system at NYU WIRELESS\cite{Xing2018gc}.}
	\label{fig:cn_sdr_TX}
	\vspace{-0.5em}
\end{figure}

Two EVBs with configuration $N = 11$; $S\langle2:0\rangle=$ ``110"; $\alpha = 1$ GHz, were interfaced with the 142 GHz channel sounder at NYU WIRELESS, replacing the bulky baseband with a single board implementation\cite{Xing2018gc}. A simplified illustration of the system is presented in Fig. \ref{fig:cn_sdr_TX}, highlighting the peripheral component interconnect-based extensions for instrumentation (PXIe) chassis developed by National Instruments Corp. used to implement the baseband. System blocks follow the conceptual block diagram presented in Fig. \ref{fig:Sl_corr}(a) and (b) with the PXIe-1085 chassis encompassing the baseband and one local oscillator (LO). 

 \begin{figure}[htbp]
	\centering
	\includegraphics[width=0.44\textwidth]{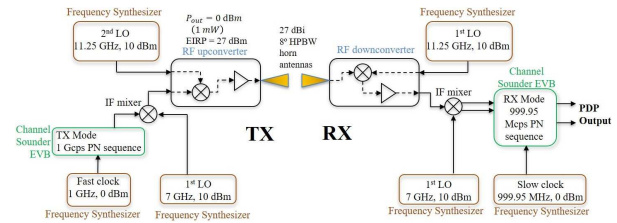}
	\caption{The 142 GHz channel sounding system TX and RX basebands replaced by the EVB and frequency synthesizers for $\alpha$ and $\beta$ in a novel baseband implementation.}
	\label{fig:cn_sdr_TXmod}
	\vspace{-0.5em}
\end{figure}
 
 The TX and RX basebands originally built on the PXIe chassis were replaced with a single EVB and frequency synthesizers, as presented in Fig. \ref{fig:cn_sdr_TXmod}. The chassis was the bulkiest system component and replacing it reduced weight by over 40 lbs and shrunk volume from 18"$\times$18"$\times$8" to 7.2"$\times$5.7"$\times$0.1". Also, eliminating the PXIe chassis and associated accessories, including cables and adapters, reduced the total cost of the system by thousands of dollars. The RX EVB integrated the critical sliding correlator on-board and thus eliminated the external correlator block allowing a novel and simple single board baseband setup.
 
 \section{Antenna XPD measurements}
Following EVB integration with the 142 GHz NYU WIRELESS channel sounder, PDPs were acquired at increasing distances up to 5 m with antennas pointed boresight. The horn antenna gain was 27 dBi and half power beamwidth (HPBW) was 8-degrees. PDPs were measured at two polarization conditions, namely, co-polarized (V-V) and cross-polarized (V-H). The V-H was achieved with a waveguide twist at the RX that rotated the RX antenna by 90-degrees. Following guidelines in \cite{xing2018vtc}, a TX and RX antenna height of 1.5 m was set such that ground reflections would not be captured by the RX antenna and measurements were made over a TX-RX separation from 3 m to 5 m in 0.5 m increments. This separation ensured operation in the far-field of the antenna. Before the measurements, calibration was completed to determine a linear range and system gain. Within the linear range, the received power at the output dropped by 10 dB for an equal 10 dB increase in signal attenuation. Operation within the linear range assured that the true received power at the RX antenna is accurately represented at the output.

\begin{figure}[htbp]
	\centering
	\includegraphics[width=0.44\textwidth]{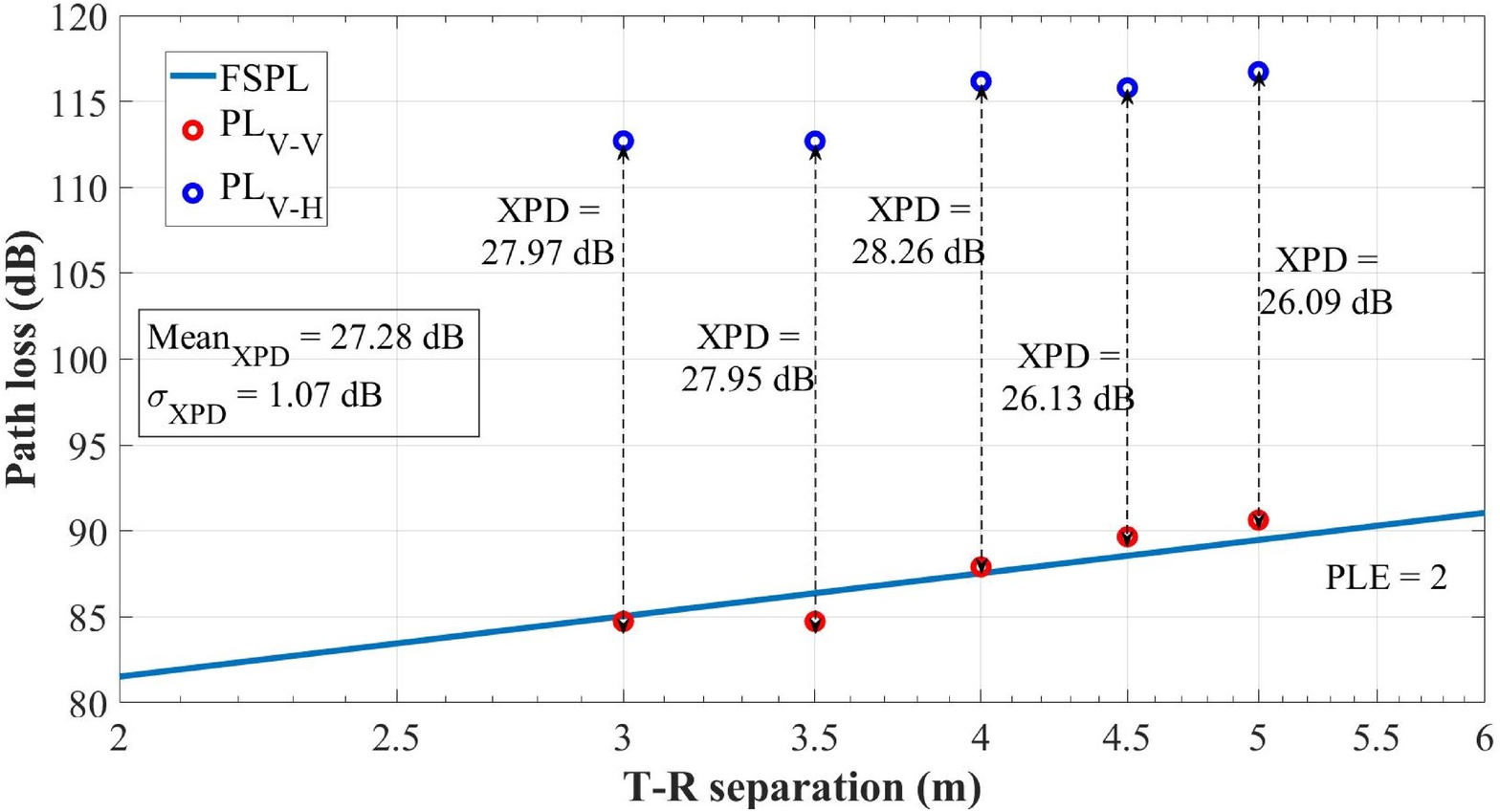}
	\caption{XPD measurement results using the EVB at 142 GHz and 27 dBi 8$^{\circ}$ HPBW horn antennas. $PL_{V-V}$ measurements follow FSPL and $PL_{V-H}$ for cross-polarized antennas show average XPD of 27.28 dB.}
	\label{fig:xpol}
	\vspace{-1.2em}
\end{figure}

For the received powers, the system gain was deducted and with the obtained true received power after the RX-antenna, path loss was calculated using \eqref{eq:pl} for both V-V and V-H configurations at each measured distance; the XPD was calculated using \eqref{eq:xpd}. Results were obtained as graphed in Fig. \ref{fig:xpol}. V-V measurements were verified with the free-space path loss (FSPL) having a path-loss exponent, $PLE=2$, and V-H measurements showed an average XPD ($\mu$) of 27.28 dB with standard deviation ($\sigma_{XPD}$) of 1.07 dB, which is 3 dB less than the $\mu$ for narrow-to-narrow antenna XPD in \cite{xing2018vtc} at 73 GHz with equal $\sigma_{XPD}$\cite{Mac2015ia}.
\vspace{-1.0em}

\color{black}     
\section*{Conclusion}
The measurement results for a programmable single-board system featuring the world's first wideband sliding correlator-based channel sounder IC fabricated in 65 nm CMOS are highlighted. The EVB offers a remarkable baseband operation at 1 Gcps, achieves a null-to-null RF bandwidth of 2 GHz, and can resolve multipath components at a nanosecond delay. The EVB can be interchangeably used at the TX or RX, configured with selection switches and offers synchronization capability. \textcolor{black}{The single-board solution potentially reduces the cost of channel sounder systems from hundreds to a few thousand dollars, simultaneously reducing system complexity, delivering higher multipath resolution, and miniaturizing the system. The XPD measurements showcase the EVB's applicability for real-world channel sounding.}

\section*{Acknowledgment}
This work is supported by the NYU WIRELESS Industrial Affiliates program and National Science Foundation grant 1909206 and 2037845. 
        
\bibliographystyle{IEEEtran}
\bibliography{references}

\begin{thebibliography}{10}
\providecommand{\url}[1]{#1}
\csname url@samestyle\endcsname
\providecommand{\newblock}{\relax}
\providecommand{\bibinfo}[2]{#2}
\providecommand{\BIBentrySTDinterwordspacing}{\spaceskip=0pt\relax}
\providecommand{\BIBentryALTinterwordstretchfactor}{4}
\providecommand{\BIBentryALTinterwordspacing}{\spaceskip=\fontdimen2\font plus
\BIBentryALTinterwordstretchfactor\fontdimen3\font minus
  \fontdimen4\font\relax}
\providecommand{\BIBforeignlanguage}[2]{{%
\expandafter\ifx\csname l@#1\endcsname\relax
\typeout{** WARNING: IEEEtran.bst: No hyphenation pattern has been}%
\typeout{** loaded for the language `#1'. Using the pattern for}%
\typeout{** the default language instead.}%
\else
\language=\csname l@#1\endcsname
\fi
#2}}
\providecommand{\BIBdecl}{\relax}
\BIBdecl

\bibitem{KO2019cm}
K.~{K.O.}, W.~{Choi}, Q.~{Zhong}, N.~{Sharma}, Y.~{Zhang}, R.~{Han},
  Z.~{Ahmad}, D.~{Kim}, S.~{Kshattry}, I.~R. {Medvedev}, D.~J. {Lary},
  H.~{Nam}, P.~{Raskin}, and I.~{Kim}, ``Opening terahertz for everyday
  applications,'' \emph{IEEE Communications Magazine}, vol.~57, no.~8, pp.
  70--76, Aug 2019.

\bibitem{Rappaport2019ia}
T.~S. {Rappaport}, Y.~{Xing}, O.~{Kanhere}, S.~{Ju}, A.~{Madanayake},
  S.~{Mandal}, A.~{Alkhateeb}, and G.~C. {Trichopoulos}, ``Wireless
  communications and applications above 100 {GHz}: Opportunities and challenges
  for {6G} and beyond,'' \emph{IEEE Access}, vol.~7, pp. 78\,729--78\,757,
  2019.

\bibitem{ghosh2019ia}
A.~Ghosh, A.~Maeder, M.~Baker, and D.~Chandramouli, ``{5G} evolution: A view on
  {5G} cellular technology beyond {3GPP} release 15,'' \emph{IEEE Access},
  vol.~7, pp. 127\,639--127\,651, Sep 2019.

\bibitem{Akyildiz2014pc}
I.~F. Akyildiz, J.~M. Jornet, and C.~Han, ``Terahertz band: Next frontier for
  wireless communications,'' \emph{Physical Communication}, vol.~12, pp.
  16--32, Sep 2014.

\bibitem{pirkl2008wc}
R.~J. Pirkl and G.~D. Durgin, ``Optimal sliding correlator channel sounder
  design,'' \emph{IEEE Transactions on Wireless Communications}, vol.~7, no.~9,
  pp. 3488--3497, Sep 2008.

\bibitem{Wu2019iscas}
T.~{Wu}, T.~S. {Rappaport}, M.~E. {Knox}, and D.~{Shahrjerdi}, ``A wideband
  sliding correlator-based channel sounder with synchronization in 65 nm
  {CMOS},'' in \emph{2019 IEEE International Symposium on Circuits and Systems
  (ISCAS)}, May 2019, pp. 1--5.

\bibitem{Mac2017sac}
G.~R. {MacCartney} and T.~S. {Rappaport}, ``A flexible millimeter-wave channel
  sounder with absolute timing,'' \emph{IEEE Journal on Selected Areas in
  Communications}, vol.~35, no.~6, pp. 1402--1418, June 2017.

\bibitem{Zhang2018icc}
Y.~Zhang, S.~Jyoti, C.~R. Anderson, D.~J. Love, N.~Michelusi, A.~Sprintson, and
  J.~V. Krogmeier, ``28-{GHz} channel measurements and modeling for suburban
  environments,'' in \emph{2018 IEEE International Conference on Communications
  (ICC)}, 2018, pp. 1--6.

\bibitem{Xing2021cl}
Y.~{Xing}, T.~{Rappaport}, and A.~{Ghosh}, ``Millimeter wave and sub-{THz}
  indoor radio propagation channel measurements, models, and comparisons in an
  office environment (invited paper),'' \emph{IEEE Communications Letters}, pp.
  1--5, Feb. 2021.

\bibitem{newhall1997wideband}
W.~G. Newhall, ``Wideband propagation measurement results, simulation models,
  and processing techniques for a sliding correlator measurement system,''
  Master's thesis, Virginia Tech, Nov 1997.

\bibitem{xing2018vtc}
Y.~Xing, O.~Kanhere, S.~Ju, T.~S. Rappaport, and G.~R. MacCartney,
  ``Verification and calibration of antenna cross-polarization discrimination
  and penetration loss for millimeter wave communications,'' in \emph{2018 IEEE
  88th Vehicular Technology Conference (VTC-Fall)}, 2018, pp. 1--6.

\bibitem{Mac2015ia}
G.~R. Maccartney, T.~S. Rappaport, S.~Sun, and S.~Deng, ``Indoor office
  wideband millimeter-wave propagation measurements and channel models at 28
  and 73 {GHz} for ultra-dense {5G} wireless networks,'' \emph{IEEE Access},
  vol.~3, pp. 2388--2424, 2015.

\bibitem{Wong2017tcas}
H.~Wong, W.~Lin, L.~Huitema, and E.~Arnaud, ``Multi-polarization reconfigurable
  antenna for wireless biomedical system,'' \emph{IEEE Transactions on
  Biomedical Circuits and Systems}, vol.~11, no.~3, pp. 652--660, 2017.

\bibitem{Shakya2020gc}
D.~{Shakya}, T.~{Wu}, and T.~S. {Rappaport}, ``A wideband sliding correlator
  based channel sounder in 65 nm {CMOS}: An evaluation board design,'' in
  \emph{GLOBECOM 2020 - 2020 IEEE Global Communications Conference}, 2020, pp.
  1--6.

\bibitem{Xing2018gc}
Y.~{Xing} and T.~S. {Rappaport}, ``Propagation measurement system and approach
  at 140 {GHz}-moving to {6G} and above {100 GHz},'' in \emph{2018 IEEE Global
  Communications Conference (GLOBECOM)}, Dec 2018, pp. 1--6.

\bibitem{zwick2005vt}
T.~Zwick, T.~J. Beukema, and H.~Nam, ``Wideband channel sounder with
  measurements and model for the 60 {GHz} indoor radio channel,'' \emph{IEEE
  transactions on Vehicular technology}, vol.~54, no.~4, pp. 1266--1277, Aug
  2005.

\bibitem{Girlando1999tcas}
G.~Girlando and G.~Palmisano, ``Noise figure and impedance matching in {RF}
  cascode amplifiers,'' \emph{IEEE Transactions on Circuits and Systems II:
  Analog and Digital Signal Processing}, vol.~46, no.~11, pp. 1388--1396, 1999.

\bibitem{Takeuchi2001pimrc}
T.~{Takeuchi} and M.~{Tamura}, ``A ultra-wide band channel sounder for mobile
  communication systems,'' in \emph{12th IEEE International Symposium on
  Personal, Indoor and Mobile Radio Communications. PIMRC 2001. Proceedings
  (Cat. No.01TH8598)}, vol.~2, Sep 2001, pp. E--E.

\end{thebibliography}

\end{document}